\documentclass[12pt,a4paper]{article}
\usepackage[utf8]{inputenc}
\usepackage{natbib}
\usepackage[margin=1in]{geometry}

\usepackage{amsthm, amssymb, amsmath, appendix, physics, url}
\usepackage{bm}
\usepackage{setspace}
\usepackage{hyperref}

\usepackage{mathtools}
\mathtoolsset{showonlyrefs=true}

\usepackage{fourier}
\usepackage[T1]{fontenc}

\DeclareMathAlphabet{\mathcal}{OMS}{cmsy}{m}{n}

\usepackage[mathlines]{lineno}
\renewcommand{\thesection}{\Roman{section}}
\renewcommand{\thesubsection}{\Alph{subsection}}

\usepackage{titlesec}

\titleformat{\section}
  {\normalsize\bfseries\centering} 
  {\thesection.} 
  {0.5em} 
  {} 

\titleformat{\subsection}
  {\normalfont\centering} 
  {\thesubsection.}    
  {0.5em}              
  {\itshape}    

\titleformat{\subsubsection}
  {\normalfont\centering} 
  {\thesubsubsection}    
  {0.5em}              
  {}    

\titleformat{name=\section, numberless}
  {\normalsize\centering\scshape}
  {}
  {0em}
  {}

\theoremstyle{definition}

\theoremstyle{plain}
\newtheorem{theorem}{Theorem}

\newtheorem{axiom}{Axiom}
\newtheorem{proposition}{Proposition}
\newtheorem{corollary}{Corollary}

\usepackage{tikz}
\usetikzlibrary{intersections, calc, arrows.meta}

\makeatletter

\@addtoreset{figure}{section}
\@addtoreset{table}{section}
\makeatother 

\newcommand{\pstar}{$^\star$}

\title{Agreement with reservation of judgment under risk}
\author{Leo Kurata and Kensei Nakamura}
\date{\today}

\begin{document}

\onehalfspacing
\sloppy

\maketitle

\begingroup
\def\thefootnote{}
\footnotetext{The authors are grateful to Kaname Miyagishima, Koichi Tadenuma, and Shohei Yanagita for their helpful comments.  
This research is financially supported by KAKENHI (Nos. 25KJ2146 \& 25KJ1298). }
\footnotetext{Kurata: Graduate School of Economics, Waseda University, Tokyo, Japan (e-mail: \protect\url{leo.kurata.ac@gmail.com}). Nakamura: Graduate School of Economics, Hitotsubashi University, Tokyo, Japan (e-mail: \protect\url{kensei.nakamura.econ@gmail.com}).}
\endgroup

\begin{abstract}
This paper studies preference aggregation under risk. In our model, each agent has an incomplete preference relation represented by a set of expected utility functions.
The classical Pareto principle is silent on agreement involving indecisiveness.
To examine the implications of respecting such agreement, we introduce the Paretian principle that can be applied when some individuals reserve their judgment. 
Our main result shows that, under this principle, for each combination of individuals' utility functions, there exists a corresponding social utility function constructed as a weighted sum of the individual ones.
These aggregation rules guarantee natural properties that the standard Pareto principle fails to ensure. 
\vspace{1mm}
\end{abstract}

\section{Introduction}

This paper considers a collective decision-making problem in which individuals are unsure of their own tastes over risky prospects, and hence sometimes reserve their judgment. 
The enormous literature on markets, mechanisms, and normative welfare criteria has examined the Pareto principle (or efficiency) as the most basic concept for comparing social states.
This criterion requires that if all individuals agree on their preference between two alternatives, then this agreement should be reflected in the social preference.
In other words, under this principle, the social preference ought to be compatible with unanimous agreement among individuals. 
Despite its popularity, this criterion overlooks a form of agreement among individuals in situations where they may withhold their judgment (i.e., when their preferences may violate completeness).

Suppose that some individuals conclude that policy A is better than policy B, but the remaining individuals reserve their judgment since it is difficult for them to compare these policies. 
In this case, the Pareto principle says nothing because there does not exist an agreement such that policy A is better than policy B. 
However, there does still exist a form of agreement among individuals: policy B is \textbf{not} weakly better than policy A.
To take these unanimous agreements into consideration, 
this paper examines a dual of the Pareto principle, which we call \textbf{\textit{the Pareto{\pstar} principle}}.%
\footnote{This axiom was also studied in \citet*{kurata2025reservation} in the context of ambiguity. For details, see Section \ref{sec:literature}.} 
This requires that if all individuals think one risky prospect is not weakly better than another, then the social preference should reach the same conclusion.

As in \citet*{danan2015harsanyi}, we study the framework in which all agents follow the expected multi-utility model \citep*{dubra2004expected}: 
Each of them has a set of utility functions, and concludes one prospect is weakly better than another 
if and only if the expected utility of the former is at least as high as that of the latter for each plausible utility function. 
That is, they reserve their judgment between two lotteries unless all utility functions provide the same evaluation for these lotteries.
\citet*{danan2015harsanyi} studied the implications of the Pareto principle in this setup. 
Roughly speaking, they showed that each social utility function must be constructed from some combination of individual utility functions via a weighted utilitarian rule \`{a} la \citet*{harsanyi1955cardinal}. 

Our main result characterizes the implications of the Pareto{\pstar} principle and shows that this principle plays a role complementary to that of the standard Pareto principle. 
Specifically, fix an arbitrary utility function from each individual's set. 
According to our theorem, the combination of chosen utility functions must be respected by the social preferences under the Pareto{\pstar} principle. 
More precisely, \textbf{for each combination of individual utility functions, there must exist a social utility function that is essentially represented by a weighted sum of them}. 
These rules prohibit the social preference from ignoring any combination of individual utility functions. 

As an immediate implication, we can see that under our rules, utility functions shared by all individuals are also adopted by the social preference.
In other words, the social planner cannot arbitrarily discard utility functions that all individuals deem plausible.
Notice that the rules in \citet*{danan2015harsanyi} do not have this property since they are allowed to discard some combination of utility functions.
Furthermore, when combined with the standard Pareto principle, the Pareto{\pstar} principle suggests that if all individuals have a common set of utility functions, then the social planner must have the same set.
Therefore, the Pareto{\pstar} principle guarantees that the social planner respects the consensus found in individual preferences.

This paper is organized as follows: 
Section \ref{sec:model} formalizes our model. 
Section \ref{sec:result} provides our main theorem and explains its implications. 
Section \ref{sec:further} examines the implications of variants of the Pareto{\pstar} principle. 
Section \ref{sec:literature} discusses the related literature.

\section{Model}\label{sec:model}

We study the model examined in \citet*{danan2015harsanyi}. 
Let $Z$ be a finite set of outcomes. 
We denote by $\Delta(Z)$ the set of lotteries over $Z$. 
Let $N = \{1,2,\ldots, n\}$ be the set of individuals with $n\geq 2$. The social planner is denoted by $0$. 
Each agent $i\in N\cup\{0\}$ has a nondegenerate preference relation $\succsim_i$ over $\Delta(Z)$.
For all $l,l'\in \Delta(Z)$, we write $l\succsim_i l'$ if agent $i$ thinks that $l$ is at least as desirable as $l'$. 
The symmetric and asymmetric parts of $\succsim_i$ are denoted by $\sim_i$ and $\succ_i$, respectively. 

We assume that each relation $\succsim_i$ is represented by the \textit{expected multi-utility model} \citep*{dubra2004expected}.
That is, for each $i\in N\cup\{ 0\}$, there exists a nonempty compact convex set $\mathcal{U}_i$ of nonconstant functions from $Z$ to $\mathbb{R}$ such that for all $l, l' \in \Delta(Z)$, 
\begin{equation*}
    l \succsim_i l' \iff \qty[ ~ \sum_{z\in Z} l (z) u_i (z) \geq  \sum_{z\in Z} l' (z) u_i (z) ~~~\text{for all } u_i \in \mathcal{U}_i~ ]. 
\end{equation*}
That is, agent $i$ weakly prefers $l$ to $l'$ if the expected utility level of $l$ is no less than that of $l'$ for all utility functions in mind.  
Preferences in this class are not complete in general. 
Agent $i$ reserves a judgment in the comparison between $l$ and $l'$ if two utility functions in $\mathcal{U}_i$ give conflicting evaluations; that is, there exists $u_i, u'_i \in\mathcal{U}_i$ such that $\sum_{z\in Z} l (z) u_i (z) > \sum_{z\in Z} l' (z) u_i (z)$ but  $\sum_{z\in Z} l' (z) u'_i (z) > \sum_{z\in Z} l (z) u'_i (z)$. 
Note that if the set $\mathcal{U}_i$ is a singleton, then $\succsim_i$ becomes an expected utility preference relation. 

For two nonconstant functions $u$ and $u'$ from $Z$ to $\mathbb{R}$, we write $u \approx u'$ if $u = \alpha u' + \beta$ for some $(\alpha, \beta) \in \mathbb{R}_{++} \times \mathbb{R}$. 
For two nonempty compact convex sets $\mathcal{U} $ and $\mathcal{U}'$ of nonconstant functions from $Z$ to $\mathbb{R}$, we write $\mathcal{U}\approx \mathcal{U}'$ if (i) for all $u\in \mathcal{U}$, there exists $u'\in \mathcal{U}'$ such that $u\approx u'$ and (ii) for all $v'\in \mathcal{U}'$, there exists $v\in \mathcal{U}$ such that $v\approx v'$. Thus, $\mathcal{U}\approx \mathcal{U}'$ means that the difference between the two sets comes only from positive affine transformations.  

We introduce an assumption for the preference profile. We say that $(\mathcal{U}_i)_{i\in N}$ has a \textit{no-conflict pair} if there exist $z^*, z_* \in Z$ such that $u(z^*) > u(z_*)$ for all $u \in \bigcup_{i\in N}\mathcal{U}_i$. 
That is, this assumption requires that there exists a pair of outcomes such that one is obviously better than the other. 

\section{Main Characterization}\label{sec:result}

In preference aggregation, social evaluations are usually required to respect unanimous individual agreements.
The following is the most standard formalization.

\begin{axiom}[Pareto]
    For all $l,l'\in\Delta (Z)$, if $l \succsim_i l'$ for all $i\in N$, then $l \succsim_0 l'$. 
\end{axiom}

We examine the ``dual'' of the above. 
Consider a case in which some individuals strictly prefer a lottery $l'$ to another $l$, while the remaining individuals cannot compare them. 
Then, Pareto does not suggest anything because its prerequisite does not hold. 
Nevertheless, we can observe that these individuals reach another type of agreement in the sense that 
\textit{no} individual thinks $l$ to be weakly better than $l'$. 
The following axiom requires the social planner to respect such unanimity.
That is, if individuals think $l$ to be not weakly better than another $l'$, then the social planner should reach the same conclusion.

\begin{axiom}[Pareto{\pstar}]
    For all $l,l'\in\Delta (Z)$, if $l \not\succsim_i l'$ for all $i\in N$, then $l \not\succsim_0 l'$. 
\end{axiom}

According to \citet*{danan2015harsanyi}, Pareto implies that each utility function in $\mathcal{U}_0$ must be represented as a weighted sum of individuals utility functions; that is, for each $u_0 \in \mathcal{U}_0$, there exists a combination $(u_1, u_2,\ldots, u_n) \in \Pi_{i\in N} ~\mathcal{U}_i$ such that 
\begin{equation}
    u_0 =  \sum_{i\in N} \alpha_i u_i + \beta
\end{equation}
for some $(\alpha, \beta) \in \qty( \mathbb{R}^n_+\setminus \{ \mathbf{0}\} ) \times \mathbb{R}$.%
\footnote{Let $\mathbf{0}$ and $\mathbf{1}$ denote the vectors of zeros and ones in $\mathbb{R}^k$, respectively. We suppress the dependence on $k$ for simplicity.}
This result is a natural extension of \citeauthor{harsanyi1955cardinal}'s (\citeyear{harsanyi1955cardinal}) aggregation theorem to the multi-utility case.

Our main theorem shows that Pareto{\pstar} yields an implication dual to that of the original Pareto axiom. 

\begin{theorem}
\label{thm:risk}
    Suppose that each $i\in N\cup\{0 \}$ has an expected multi-utility preference relation $\succsim_i$ associated with $~\mathcal{U}_i$ and that $(\mathcal{U}_i)_{i\in N}$ has a no-conflict pair.
    Then, Pareto{\pstar} holds if and only if 
    for each $(u_1, u_2,\ldots, u_n) \in \Pi_{i\in N} ~\mathcal{U}_i$, there exists $u_0 \in \mathcal{U}_0$ such that 
    \begin{equation}
        u_0 =  \sum_{i\in N} \alpha_i u_i + \beta
    \end{equation}
    for some $(\alpha, \beta) \in \qty( \mathbb{R}^n_+\setminus \{ \mathbf{0}\} ) \times \mathbb{R}$. 
\end{theorem}

This is another extension of Harsanyi's aggregation theorem to our framework (cf.~\citealp*{de1995note}). 
According to Theorem \ref{thm:risk}, individual utility functions are aggregated in a utilitarian way for each combination $(u_1, u_2,\ldots, u_n) \in \Pi_{i\in N} ~\mathcal{U}_i$ of utility functions. 
The most important difference from the result in \citet*{danan2015harsanyi} is that Pareto{\pstar} ensures that \textbf{all combinations of utility functions are used in the aggregation process}. 
That is, these rules prohibit the social planner from ignoring any combination of individual utility functions.

\begin{proof}[\bf Proof of Theorem 1]
    ``If'' part. Let $l, l'\in \Delta(Z)$ be such that $l\not\succsim_i l'$ for all $i\in N$. Then, for each $i\in N$, there exists $u_i \in \mathcal{U}_i$ such that $\sum_{z\in Z} l'(z) u_i (z) > \sum_{z\in Z} l(z) u_i (z) $. 
    Since there exists $u_0\in \mathcal{U}_0$ such that $u_0 = \sum_{i\in N} \alpha_i u_i +\beta$ for some $(\alpha, \beta) \in \qty(\mathbb{R}_{+}^n\setminus \{\mathbf{0} \}) \times \mathbb{R}$, we have 
    \begin{equation}
        \sum_{z\in Z} l'(z) u_0 (z) = \sum_{i\in N} \alpha_i \sum_{z\in Z} l'(z) u_i (z) +\beta > \sum_{i\in N} \alpha_i \sum_{z\in Z} l(z) u_i (z) +\beta = \sum_{z\in Z} l(z) u_0 (z). 
    \end{equation}
    Therefore, we have $l\not\succsim_0 l'$. 
    
    ``Only-if'' part. 
    Suppose to the contrary that Pareto{\pstar} holds but there exists $(u_1, u_2,\ldots, u_n) \in \Pi_{i\in N} ~\mathcal{U}_i$ such that for all $u_0\in \mathcal{U}_0$, $u_0$ cannot be written as $u_0 =  \sum_{i\in N} \alpha_i u_i + \beta$ for any $(\alpha, \beta) \in \qty( \mathbb{R}^n_+ \setminus \{ \mathbf{0}\} ) \times \mathbb{R}$.
    Let $A = \qty{ \sum_{i\in N} \alpha_i u_i + \beta  ~  | ~  (\alpha, \beta) \in  \mathbb{R}^n_+  \times \mathbb{R} } \subseteq \mathbb{R}^Z$. 
    Then, since each $u_0$ is nonconstant, we have 
    \begin{equation}
    \label{eq:disjoint1}
        A \cap \mathcal{U}_0 = \emptyset. 
    \end{equation}
    Let $B = \qty{ \alpha_0 u_0 +\beta_0 \mid \qty(\alpha_0, \beta_0) \in \mathbb{R}_{++} \times \mathbb{R}, ~u_0\in \mathcal{U}_0  }\cup\{\mathbf{0}\} \subseteq \mathbb{R}^Z$. 
    Then, \eqref{eq:disjoint1} implies 
    \begin{equation}
    \label{eq:disjoint2}
        A \cap B = \{\mathbf{0}\}. 
    \end{equation}
    Note that $A$ and $B$ are convex cones.%
    \footnote{
    To see that $A$ is a cone, let $\varphi \in A$ and $\lambda\in \mathbb{R}_{++}$. Then, there exists $(\alpha, \beta) \in  \mathbb{R}^n_+  \times \mathbb{R}$ such that $\varphi = \sum_{i\in N}\alpha_i u_i + \beta$. Since $\lambda\varphi = \sum_{i\in N} \lambda\alpha_i u_i + \lambda\beta$
    also holds, $A$ is a cone. We can check that $B$ is also a cone by a similar argument. }
    By applying Theorem 2.7 of \citet*{klee1955separation}, there exists $ \lambda \in \mathbb{R}^Z \setminus \{\mathbf{0}\} $
    such that for all $\varphi \in A \setminus(-A)$ and all $ \varphi' \in B$, 
    \begin{equation}
        \sum_{z\in Z} \lambda (z) \varphi (z) > 0 \geq \sum_{z\in Z} \lambda (z) \varphi' (z). 
    \end{equation}
    We prove that $ A \cap (-A) = \{ \gamma \mathbf{1}\}_{\gamma\in \mathbb{R}}$. 
    Suppose to the contrary that there is a nonconstant element $\psi\in A \cap (-A)$. 
    Since $(\mathcal{U}_i)_{i\in N}$ has a no-conflict pair,  there exists
    $(z^*, z_*) \in Z^2$ such that for all $i\in N\cup\{0\}$ and all $u_i\in \mathcal{U}_i$, $u_i (z^*) > u_i(z_*)$. 
    Since $\psi$ is nonconstant, there is $(\alpha,\beta) \in ( \mathbb{R}^n_{+} \backslash \{ \mathbf{0}\} ) \times \mathbb{R}$ such that $\psi = \sum_{i\in N} \alpha_i u_i + \beta$. 
    Therefore, we have  $\psi(z^*) > \psi(z_*)$. 
    Similarly, $\psi \in (-A)$ implies $\psi(z^*) < \psi(z_*)$, which is a contradiction.

    Therefore, by the definitions of $A$ and $B$, for all $i\in N$, all $u_0\in \mathcal{U}_0$, all $\qty(\alpha_0 , \beta_0) \in \mathbb{R}_{++}\times \mathbb{R}$,  and all $(\alpha , \beta ) \in \mathbb{R}^n_{++} \times  \mathbb{R}^n$, 
    \begin{equation}
    \label{eq:separate}
        \sum_{z\in Z} \lambda (z) \qty( \alpha_i u_i(z) +\beta_i ) > 0 \geq \sum_{z\in Z} \lambda (z) \qty(\alpha_0 u_0 (z) + \beta_0). 
    \end{equation}
    
    For each $u \in \bigcup_{i\in N\cup \{0\}} \mathcal{U}_i$, let $\hat{u}$ denote a function such that $\sum_{z\in Z} \hat{u}(z) = 0$ and  $\hat{u} = \alpha u +\beta$ for some $(\alpha , \beta) \in \mathbb{R}_{++} \times \mathbb{R}$. 
    Then, by \eqref{eq:separate}, 
    for all $i\in N$ and all $u_0\in \mathcal{U}_0$, 
    \begin{equation}
    \label{eq:separate2}
        \sum_{z\in Z} \lambda (z)  \hat{u}_i(z)  > 0 \geq \sum_{z\in Z} \lambda (z)  \hat{u}_0 (z). 
    \end{equation}
    Let $\underline{\lambda} = - \min_{z\in Z} \lambda(z) + 1$. 
    Since $\sum_{z\in Z} \hat{u}(z) = 0$ for all $u\in \{ u_1,\ldots, u_n\} \cup\mathcal{U}_0$, we have that  for all $i\in N$ and all $u_0\in \mathcal{U}_0$,  
    \begin{equation}
    \label{eq:separate3}
        \sum_{z\in Z} \qty(\lambda (z) + \underline{\lambda})  \hat{u}_i (z) > 0 \geq \sum_{z\in Z} \qty(\lambda (z) +\underline{\lambda}) \hat{u}_0 (z). 
    \end{equation}
    Let $l , l'\in \Delta(Z)$ be such that for all $z \in Z$,
    \begin{equation}
        l(z) = {\lambda(z) + \underline{\lambda}  \over \sum_{z'\in Z} \lambda (z') + |Z|\underline{\lambda}  } ~ (>0) ~~~ \text{and} ~~~ l'(z) =  {1\over |Z|}.
    \end{equation}
    By \eqref{eq:separate3} and the definitions of $l$ and $l'$, for all $i\in N$ and all $u_0 \in \mathcal{U}_0$, 
     \begin{equation}
        \sum_{z\in Z} l (z)  \hat{u}_i (z) > \sum_{z\in Z} l' (z)  \hat{u}_i (z) ~~~\text{and} ~~~ \sum_{z\in Z} l' (z)  \hat{u}_0 (z) \geq \sum_{z\in Z} l (z) \hat{u}_0 (z). 
    \end{equation}
    By the construction of $\hat{u}$, we have $l' \not\succsim_i l$ for each $i\in N$ and $l' \succsim_0 l$. This is a contradiction to Pareto{\pstar}.  
\end{proof}

Notice that in the aggregation rules characterized by \citet*{danan2015harsanyi}, an important property ensured by Harsanyi's result is discarded. 
To see this, suppose that all individuals have expected utility preferences associated with some common function $u$ and the planner also obey expected utility theory. Then, according to Harsanyi's theorem, the planner's utility function must be $u$. 
If we extend this observation to the multi-utility case, we expect that (i) if all individuals have $u$ in their utility sets, the planner should do so, and (ii) if all individuals have a common utility set (up to positive affine transformation), then the planner's utility set should coincide with it. 
Nevertheless, since the aggregation rules in \citet*{danan2015harsanyi} allows the planner to hold any sufficiently small set of utility functions,
they do not necessarily satisfy the above intuitive and  normatively appealing properties. 

In contrast, these properties can be ensured by imposing Pareto{\pstar}. 
The following corollary formally states these implications. 

\begin{corollary}
\label{cor:una}
    Suppose that each $i\in N\cup\{0 \}$ has an expected multi-utility preference relation associated with $~\mathcal{U}_i$ and $(\succsim_i)_{i\in N}$ has a no-conflict pair.
    The following two statements hold:
    \begin{enumerate}
        \item Let $u$ be a nonconstant function from $Z$ to $\mathbb{R}$. Pareto{\pstar} implies that if there exists $u_i\in \mathcal{U}_i$ with $u_i\approx u$ for each $i\in N$, then $u_0 \approx u$ for some $u_0\in \mathcal{U}_0$. 
        \item Let $\mathcal{U}$ be a nonempty compact convex sets of nonconstant functions from $Z$ to $\mathbb{R}$. Pareto and Pareto{\pstar} imply that if $ ~\mathcal{U}_i\approx  \mathcal{U}$ for each $i\in N$, then $ \mathcal{U}_0 \approx  \mathcal{U}$. 
    \end{enumerate}
\end{corollary}

\section{Further Results}\label{sec:further}

\subsection{Non-Reversal}

We study a minimal condition for respecting individual preferences and explore its implications.
The following requires that if all individuals  prefer one lottery $l$ to another $l'$, then the social planner should not prefer $l'$ to $l$; that is, preference reversals should not occur. 

\begin{axiom}[Non-Reversal]
    For all $l,l'\in\Delta (Z)$, if $l \succ_i l'$ for all $i\in N$, then $l'\not \succsim_0 l$. 
\end{axiom}

Note that Non-Reversal is weaker than  Pareto{\pstar} and the Paretian condition for the strict part (i.e., for all $l,l'\in\Delta (Z)$, if $l \succ_i l'$ for all $i\in N$, then $l  \succ_0 l'$). 
To characterize the implication of this requirement, we focus on multi-utility representations with some additional constraints. 
For a binary relation $\succsim$ over $\Delta(Z)$, 
a set $\mathcal{U}$ of functions from $Z$ to $\mathbb{R}$ is \textit{strictly increasing}
if, for all $u\in \mathcal{U}$ and all $l, l' \in \Delta(Z)$, $l\succ l'$ implies $\sum_{z\in Z} l (z) u (z) > \sum_{z\in Z} l' (z) u (z)$. 
If we endow the set $Z$ with the discrete topology, we can without loss of generality assume that each set $\mathcal{U}$ associating an expected multi-utility preference is strictly increasing.%
\footnote{
Suppose that $\succsim$ is an expected multi-utility preference associated by $\mathcal{U}$. 
By Proposition 2 of \citet*{dubra2004expected}, there exists a function $u^*:Z\to \mathbb{R}$ such that for all $l, l' \in \Delta(Z)$, $l\succ l'$ (resp.~$l\sim l'$) implies $\sum_{z\in Z} l (z) u^* (z) >  \sum_{z\in Z} l' (z) u^* (z)$ (resp.~$\sum_{z\in Z} l (z) u^* (z) = \sum_{z\in Z} l' (z) u^* (z)$). 
Let $\mathcal{U}^* =\{ u + \varepsilon u^* \mid (\varepsilon, u)\in \mathbb{R}_{++} \times \mathcal{U}\}$, which is a strictly increasing set.  
Then, $\succsim$ is also represented by $\mathcal{U}^*$. 
To see this, let $l, l'\in \Delta(Z)$. 
If $l\succsim l'$, then 
$\sum_{z\in Z} l (z) [u(z) +\varepsilon  u^* (z)] \geq \sum_{z\in Z} l' (z)  [u(z) + \varepsilon u^* (z)]$ for all $(\varepsilon, u)\in \mathbb{R}_{++} \times \mathcal{U}$. 
For the converse, suppose that $\sum_{z\in Z} l (z) [u(z) +\varepsilon  u^* (z)] \geq \sum_{z\in Z} l' (z)  [u(z) + \varepsilon u^* (z)]$ for all $(\varepsilon, u)\in \mathbb{R}_{++} \times \mathcal{U}$.
Then, we have $\sum_{z\in Z} l (z) u(z) \geq \sum_{z\in Z} l' (z)  u(z) $ for all $u\in \mathcal{U}$, which implies $l\succsim l'$. 
The argument here mimics the proof of Corollary 1 in \citet*{evren2011multi}. 
}

The following proposition characterizes the implication of Non-Reversal.

\begin{proposition}
    Suppose that each $i\in N\cup\{0 \}$ has an expected multi-utility preference relation $\succsim_i$ associated with a strictly increasing set $\mathcal{U}_i$ and that $(\mathcal{U}_i)_{i\in N}$ has a no-conflict pair. 
    Then, Non-Reversal holds if and only if there exists
    $(u_0, u_1,\ldots, u_n) \in \Pi_{i\in N\cup\{ 0 \}} ~\mathcal{U}_i$ such that 
    \begin{equation}
        u_0 =  \sum_{i\in N} \alpha_i u_i + \beta
    \end{equation}
    for some $(\alpha, \beta) \in ( \mathbb{R}^n_+\setminus \{ \mathbf{0}\} ) \times \mathbb{R}$. 
\end{proposition}

This proposition shows that under Non-Reversal, there is at least one social utility function $u_0\in \mathcal{U}_0$ and at least one combination $(u_1, u_2, \ldots, u_n) \in \Pi_{i\in N} ~\mathcal{U}_i$
such that $u_0$ is constructed as a weighted sum of $(u_1, u_2, \ldots, u_n)$. 
If this property holds for all social utility functions, the rule corresponds to the implication of Pareto. 
On the other hand, if this property holds for all combinations of individual utility functions, it corresponds to the implication of Pareto{\pstar}. 

\subsection{Bi-utilitarianism}

\citet*{danan2015harsanyi} examined an axiom weaker than  Pareto, which is defined as follows. 

\begin{axiom}[Pareto Indifference]
    For all $l,l'\in\Delta (Z)$, if $l \sim_i l'$ for all $i\in N$, then $l \sim_0 l'$. 
\end{axiom}

They showed that Pareto Indifference holds if and only if the social preference takes a form of \textit{bi-utilitarianism}, that is, for all $u_0\in \mathcal{U}_0$, there exists $(u_i, v_i)_{i\in N} \in \prod_{i\in N} \mathcal{U}_i^2$ such that $u_0 = \sum_{i\in N} \alpha_i u_i - \sum_{i\in N}\alpha'_i v_i +\beta$ for some $(\alpha, \alpha', \beta) \in  \mathbb{R}^n_+ \times \mathbb{R}^n_+ \times \mathbb{R}$ with $(\alpha, \alpha') \neq (\mathbf{0}, \mathbf{0})$. 
The part with negative weights appears under Pareto Indifference because this axiom allows the social preference to respond to individuals' ones in the opposite direction (e.g., $l \succsim_i l'$ for all $i\in N$ but $l' \succsim_0 l$). 

Similarly, we introduce a weakening of Pareto{\pstar} and examine its implications.
For all $i\in N$ and all $l, l' \in \Delta (Z)$, we write $l \bowtie_i l'$ if $l\not\succsim_i l'$ and $l' \not\succsim_i l$. 
Thus, $l \bowtie_il'$ means that agent $i$ cannot compare $l$ and $l'$. 
The following axiom requires that if all individuals cannot compare $l$ and $l'$, then the social planner should also reserve judgment. 

\begin{axiom}[Pareto Incomparability]
    For all $l,l'\in\Delta (Z)$, if $l \bowtie_i l'$ for all $i\in N$, then $l \bowtie_0 l'$. 
\end{axiom}

We say that a combination $(u_i, v_i)_{i\in N} \in \prod_{i\in N} \mathcal{U}_i^2 $ is \textit{bi-independent} if $\sum_{i\in N} \alpha_i u_i + \beta \neq \sum_{i\in N} \alpha'_i v_i + \beta'$ for all $(\alpha, \beta), (\alpha', \beta') \in  \mathbb{R}^n_+ \times \mathbb{R}$ with $(\alpha, \alpha') \neq (\mathbf{0}, \mathbf{0})$.
This means that $\{ u_1, \ldots, u_n, v_1,\ldots,v_n,\mathbf{1}\}$ is linearly independent (note that to satisfy this condition, $|Z| \geq 2n+ 1$ must hold). 
The following proposition shows that Pareto Incomparability implies that each combination $(u_i, v_i)_{i\in N} \in \prod_{i\in N} \mathcal{U}_i^2 $ is respected by the social planner through the bi-utilitarian aggregation.
Note that the converse does not hold. 

\begin{proposition}
    Suppose that each $i\in N\cup\{0 \}$ has an expected multi-utility preference relation $\succsim_i$ associated with $\mathcal{U}_i$. 
    Then, Pareto Incomparability implies that for each bi-independent combination $(u_i, v_i)_{i\in N} \in \prod_{i\in N} \mathcal{U}_i^2 $, there exists $u_0\in \mathcal{U}_0$ such that 
    \begin{equation}
        u_0 =  \sum_{i\in N} \alpha_i u_i - \sum_{i\in N} \alpha'_i v_i + \beta
    \end{equation}
    for some $(\alpha, \alpha', \beta) \in  \mathbb{R}^n_+ \times \mathbb{R}^n_+ \times \mathbb{R}$ with $(\alpha, \alpha') \neq (\mathbf{0}, \mathbf{0})$. 
\end{proposition}

\section{Related Literature}\label{sec:literature}

We finally discuss the related literature. 
\citet*{harsanyi1955cardinal} is the seminal work on preference aggregation under risk. 
Harsanyi showed that if all individuals and the social planner have expected utility preferences, then the Pareto principle yields a utilitarian social evaluation. 
This aggregation theorem evoked enormous discussions on its ethical meaning (e.g., \citealp*{Sen1977,Sen1986,Weymark1991,fleurbaey2016utilitarian}) and distributional equity (e.g., \citealp*{diamond1965,epstein1992quadratic,fleurbaey2010assessing}). 

The aggregation problem where agents have incomplete multi-utility preferences was considered in \citet*{danan2013aggregating,danan2015harsanyi}. 
Since the former studies a multi-profile setup, the latter is more closely related to this paper. 
\citet*{danan2015harsanyi} characterized the implications of the Pareto principle and a unanimity principle for indifference relations. 
Our paper, on the other hand, studies the dual axiom and shows that it has a complementary implication (Theorem \ref{thm:risk}). 
The aggregation rules characterized in our theorem can ensure natural unanimity properties 
that are overlooked by those of \citet{danan2015harsanyi} (Corollary \ref{cor:una}). 

Under ambiguity, \citet*{danan2016robust} and \citet*{pivato2024bayesian} studied the setting
in which agents have incomplete preference relations with multiple priors \citep*{bewley2002knightian}. 
\citet*{danan2016robust} studied the implications of the Pareto principle and showed a partial possibility result. 
Its limitation stems from spurious unanimity, which motivates them to introduce a weaker axiom. 
Their second result shows that under the weaker Pareto principle, the social belief set is included in the convex hull of the union of individual sets. 
Equivalently, each social prior must be constructed from some combination of individual priors by linear pooling.
\citet*{pivato2024bayesian} provided another foundation for these belief aggregation rules in a different domain. 
These rules correspond to the aggregation rules derived from the Pareto principle under risk \citep*{danan2015harsanyi}. 

In the companion paper, \citet*{kurata2025reservation}, we introduced the Pareto{\pstar} principle in the framework with ambiguity. 
Since this axiom inherits the problem of the original Pareto principle, it does not yield a positive result. 
\citet*{kurata2025reservation} also examined a weakening of the Pareto{\pstar} principle and characterized belief aggregation rules in which each combination of individual priors is respected by some social prior.
This result corresponds to the main theorem of this paper. 
Despite this similarity, the proof strategy used there cannot be adapted to our characterization; hence, this paper also makes a technical contribution to the literature.
In the proof of \citet*{kurata2025reservation}, the separating hyperplane theorem is directly applicable to the social belief set and the convex hull of some combination of individual priors, and the coefficient of the hyperplane corresponds to the ambiguous prospects that lead to a contradiction. 
On the other hand, we consider the cones generated from utility functions and then apply a variant of the separating hyperplane theorem to the extended sets. 
The coefficient of a hyperplane does not necessarily correspond to a lottery because the sum of the coefficients may not equal 1 and each element may be negative. 
We deal with this issue by offering an adjustment method. 

\section*{APPENDIX}
\renewcommand{\thesubsection}{A.\arabic{subsection}}
\setcounter{subsection}{0}

\subsection{Proof of Proposition 1}

``If'' part. Let $l, l'\in \Delta(Z)$ be such that $l \succ_i l'$ for all $i\in N$. Then, for each $i\in N$, Since $\mathcal{U}_i$ is a strictly increasing set, we have $\sum_{z\in Z} l'(z) u_i (z) > \sum_{z\in Z} l(z) u_i (z) $ for all $u_i \in \mathcal{U}_i$. 
    Since there exists $(u_0, u_1,\ldots, u_n) \in \Pi_{i\in N\cup\{ 0 \}} ~\mathcal{U}_i$ such that $u_0 = \sum_{i\in N} \alpha_i u_i +\beta$ for some $(\alpha, \beta) \in \qty(\mathbb{R}_{+}^n\setminus \{\mathbf{0} \}) \times \mathbb{R}$, we have 
    \begin{equation}
        \sum_{z\in Z} l'(z) u_0 (z) = \sum_{i\in N} \alpha_i \sum_{z\in Z} l'(z) u_i (z) +\beta > \sum_{i\in N} \alpha_i \sum_{z\in Z} l(z) u_i (z) +\beta = \sum_{z\in Z} l(z) u_0 (z). 
    \end{equation}
    Therefore, we have $l\not\succsim_0 l'$.

    ``Only-if'' part. 
    Suppose to the contrary that Non-Reversal holds but, for all $(u_0, u_1,\ldots, u_n) \in \Pi_{i\in N\cup\{ 0 \}} ~\mathcal{U}_i$, $u_0$ cannot be written as $u_0 =  \sum_{i\in N} \alpha_i u_i + \beta$ 
    for any $(\alpha, \beta) \in \qty( \mathbb{R}^n_+\setminus \{ \mathbf{0}\} ) \times \mathbb{R}$.
    Let $A  \subseteq \mathbb{R}^Z $ denote the set $\qty{ \sum_{i\in N} \alpha_i u_i + \beta  ~  | ~  (\alpha, \beta ) \in  \mathbb{R}^n_+  \times \mathbb{R}, ~(u_1,\ldots, u_n) \in  \prod_{i\in N} ~\mathcal{U}_i }$. 
    Then, since each $u_0 \in \mathcal{U}_0$ is nonconstant, we have 
    \begin{equation}
    \label{eq:disjoint1_nr}
        A \cap \mathcal{U}_0 = \emptyset. 
    \end{equation}
    Let $B = \{ \alpha_0 u_0 +\beta_0 \mid (\alpha_0, \beta_0) \in \mathbb{R}_{++} \times \mathbb{R}, ~ u_0\in \mathcal{U}_0  \}\cup\{\mathbf{0}\}  \subseteq \mathbb{R}^Z$. 
    Then, \eqref{eq:disjoint1_nr} implies $ A \cap B = \{\mathbf{0}\}$. 
    Note that $A$ and $B$ are convex cones.
    By applying Theorem 2.7 of \citet*{klee1955separation}, there exists $ \lambda \in \mathbb{R}^Z \setminus \{\mathbf{0}\} $
    such that for all $\varphi \in A \setminus(-A)$ and $ \varphi' \in B$, 
    \begin{equation}
        \sum_{z\in Z} \lambda (z) \varphi (z) > 0 \geq \sum_{z\in Z} \lambda (z) \varphi' (z). 
    \end{equation}
    By the argument in the proof of Theorem \ref{thm:risk}, we have $ A \cap (-A) = \{ \gamma \mathbf{1}\}_{\gamma\in \mathbb{R}}$.
    
    Therefore, by the definitions of $A$ and $B$, for all $(u_0, u_1,\ldots, u_n) \in \Pi_{i\in N\cup\{ 0 \}} ~\mathcal{U}_i$, all $(\alpha_0 , \beta_0) \in \mathbb{R}_{++}\times \mathbb{R}$,  and all $(\alpha , \beta ) \in \mathbb{R}^n_{++} \times  \mathbb{R}^n$, 
    \begin{equation}
    \label{eq:separate_nr}
        \sum_{z\in Z} \lambda (z) \qty( \alpha_i u_i(z) +\beta_i ) > 0 \geq \sum_{z\in Z} \lambda (z) \qty(\alpha_0 u_0 (z) + \beta_0). 
    \end{equation}

    For each $u \in \bigcup_{i\in N\cup \{0\}} \mathcal{U}_i$, let $\hat{u}$ denote a function such that $\sum_{z\in Z} \hat{u}(z) = 0$ and  $\hat{u} = \alpha u +\beta$ for some $(\alpha , \beta) \in \mathbb{R}_{++} \times \mathbb{R}$. 
    Then, by \eqref{eq:separate_nr}, 
    for all $(u_0, u_1,\ldots, u_n) \in \Pi_{i\in N\cup\{ 0 \}} ~\mathcal{U}_i$, 
    \begin{equation}
    \label{eq:separate2_nr}
        \sum_{z\in Z} \lambda (z)  \hat{u}_i(z)  > 0 \geq \sum_{z\in Z} \lambda (z)  \hat{u}_0 (z). 
    \end{equation}
    Let $\underline{\lambda} = - \min_{z\in Z} \lambda(z) + 1$. 
    Since $\sum_{z\in Z} \hat{u}(z) = 0$ for all $u \in \bigcup_{i\in N\cup \{0\}} \mathcal{U}_i$, we have that  for all $i\in N$ and all $u_0\in \mathcal{U}_0$,
    \begin{equation}
    \label{eq:separate3_nr}
        \sum_{z\in Z} \qty(\lambda (z) + \underline{\lambda})  \hat{u}_i (z) > 0 \geq \sum_{z\in Z} \qty(\lambda (z) +\underline{\lambda}) \hat{u}_0 (z). 
    \end{equation}
    Let $l, l' \in \Delta(Z)$ be such that for all $z \in Z$,
    \begin{equation}
        l(z) = {\lambda(z) + \underline{\lambda}  \over \sum_{z'\in Z} \lambda (z') + |Z|\underline{\lambda}  } ~ (>0) ~~~\text{and}~~~ l'(z) =  {1\over |Z|}.
    \end{equation}
    By \eqref{eq:separate3_nr} and the definitions of $l$ and $l'$, for all $i\in N$ and all $u_0 \in \mathcal{U}_0$, 
     \begin{equation}
        \sum_{z\in Z} l (z)  \hat{u}_i (z) > \sum_{z\in Z} l' (z)  \hat{u}_i (z) ~~~\text{and} ~~~ \sum_{z\in Z} l' (z)  \hat{u}_0 (z) \geq \sum_{z\in Z} l (z) \hat{u}_0 (z). 
    \end{equation}
    By the construction of $\hat{u}$, we have $l \succ_i l'$ for each $i\in N$ and $l' \succsim_0 l$. This is a contradiction to Non-Reversal.   
    \qed

\subsection{Proof of Proposition 2}

Suppose to the contrary that Pareto Incomparability holds but there exists a bi-independent combination $(u_i, v_i)_{i\in N} \in \prod_{i\in N} \mathcal{U}_i^2 $ such that for all $u_0\in \mathcal{U}_0$, $u_0$ cannot be written as $u_0 =  \sum_{i\in N} \alpha_i u_i - \sum_{i\in N} \alpha'_i v_i + \beta$ for any $(\alpha, \alpha', \beta) \in  \mathbb{R}^n_+ \times \mathbb{R}^n_+ \times \mathbb{R}$ with $(\alpha, \alpha') \neq (\mathbf{0}, \mathbf{0})$. 
Let $A  \subseteq \mathbb{R}^Z$ denote the set $\qty{ \sum_{i\in N} \alpha_i u_i - \sum_{i\in N} \alpha'_i v_i + \beta ~  | ~  (\alpha, \alpha', \beta) \in  \mathbb{R}^n_+ \times \mathbb{R}^n_+ \times \mathbb{R} }$. 
Then, since each $u_0$ is nonconstant, we have 
\begin{equation}
\label{eq:disjoint1_inc}
    A \cap \mathcal{U}_0 = \emptyset. 
\end{equation}
Let $B = \qty{ \alpha_0 u_0 +\beta_0 \mid (\alpha_0, \beta_0) \in \mathbb{R}_{++} \times \mathbb{R}, ~u_0\in \mathcal{U}_0  }\cup\{\mathbf{0}\}  \subseteq \mathbb{R}^Z$. 
Then, \eqref{eq:disjoint1_inc} implies $ A \cap B = \{\mathbf{0}\}$. 
Note that $A$ and $B$ are convex cones.
By applying Theorem 2.7 of \citet*{klee1955separation}, there exists $ \lambda \in \mathbb{R}^Z \setminus \{\mathbf{0}\} $
such that for all $\varphi \in A \setminus(-A)$ and all $ \varphi' \in B$, 
\begin{equation}
\sum_{z\in Z} \lambda (z) \varphi (z) > 0 \geq \sum_{z\in Z} \lambda (z) \varphi' (z). 
\end{equation} 

Since $(u_i, v_i)_{i\in N} \in \prod_{i\in N} \mathcal{U}_i^2 $ is bi-independent, $u_i \notin A \cap (-A)$ and  $v_i \notin A \cap (-A)$ for each $i\in N$. 
By the definitions of $A$ and $B$, for all $i\in N$, all $u_0\in \mathcal{U}_0$, all $(\alpha_0 , \beta_0) \in \mathbb{R}_{++}\times \mathbb{R}$,  and all $(\alpha , \beta ), (\alpha' , \beta') \in \mathbb{R}^n_{++} \times  \mathbb{R}^n$, 
\begin{equation}
    \sum_{z\in Z} \lambda (z) ( \alpha_i u_i(z) +\beta_i ) > 0 > \sum_{z\in Z} \lambda (z) \qty( \alpha'_i v_i(z) +\beta'_i )
\end{equation}
and 
\begin{equation}
    \sum_{z\in Z} \lambda (z) \qty(\alpha_0 u_0 (z) + \beta_0) \leq 0. 
\end{equation}

For each $u\in \{ u_1,\ldots, u_n, v_1,\ldots,v_n\}\cup\mathcal{U}_0$, we define $\hat{u}$ in a way similar to the proof of Theorem \ref{thm:risk}. 
Then, 
for all $i\in N$ and all $u_0\in \mathcal{U}_0$, 
\begin{equation}
\label{eq:separate2}
    \sum_{z\in Z} \lambda (z)  \hat{u}_i(z)  > 0 > \sum_{z\in Z} \lambda (z)  \hat{v}_i (z)~~~~\text{and}~~~~\sum_{z\in Z} \lambda (z) \hat{u}_0 \leq 0. 
\end{equation}
    Let $\underline{\lambda} = - \min_{z\in Z} \lambda(z) + 1$. 
    Since $\sum_{z\in Z} \hat{u}(z) = 0$ for all $u\in \{ u_1,\ldots, u_n, v_1,\ldots,v_n\} \cup\mathcal{U}_0$, we have that  for all $i\in N$ and all $u_0\in \mathcal{U}_0$,  
    \begin{equation}
    \label{eq:separate3_inc}
        \sum_{z\in Z} \qty(\lambda (z)  + \underline{\lambda}) \hat{u}_i(z)  > 0 > \sum_{z\in Z}  \qty(\lambda (z)  + \underline{\lambda})  \hat{v}_i (z)~~~~\text{and}~~~~\sum_{z\in Z}  \qty(\lambda (z)  + \underline{\lambda}) \hat{u}_0 \leq 0.
    \end{equation}
    Let $l, l' \in \Delta(Z)$ be such that for all $z\in Z$,  
    \begin{equation}
        l(z) = {\lambda(z) + \underline{\lambda}  \over \sum_{z'\in Z} \lambda (z') + |Z|\underline{\lambda}  } ~ (>0) ~~~ \text{and} ~~~ l'(z) =  {1\over |Z|}. 
    \end{equation}
    By \eqref{eq:separate3_inc} and the definitions of $l$ and $l'$, for all $i\in N$, 
     \begin{equation}
        \sum_{z\in Z} l (z)  \hat{u}_i (z) > \sum_{z\in Z} l' (z)  \hat{u}_i (z) ~~~\text{and} ~~~ \sum_{z\in Z} l' (z)  \hat{u}_i (z) > \sum_{z\in Z} l (z) \hat{u}_i (z),
    \end{equation}
     but for all $u_0 \in \mathcal{U}_0$, 
    \begin{equation}
         \sum_{z\in Z} l (z)  \hat{u}_0 (z) \leq \sum_{z\in Z} l' (z) \hat{u}_0 (z). 
    \end{equation}
    By the construction of $\hat{u}$, we have $l' \bowtie_i l$ for each $i\in N$ and $l' \succsim_0 l$. This is a contradiction to  Pareto Incomparability. 
    \qed

%%%%%%%%%%%%%%%%%%%%%%%%
%%%%%%%%%%%%%%%%%%%%%%%
\bibliographystyle{econ-aea}
\bibliography{reference}
%%%%%%%%%%%%%%%%%%%%%%%%%%%%%%%
%%%%%%%%%%%%%%%%%%%%%%%%%%%%%%%

\end{document}